%% file: 01_dts.tex
\documentclass[runningheads]{llncs}
\usepackage[
    left = `,% 
    right = ',% 
    leftsub = ``,% 
    rightsub = '' %
]{dirtytalk}
\usepackage{longtable}
\usepackage{multirow}
\usepackage{subcaption}
\usepackage{tabularx}   
\usepackage{url}  

\usepackage{graphicx}
\usepackage{hyperref}
\hypersetup{
    colorlinks=true,
    linkcolor=blue,
    filecolor=magenta,      
    urlcolor=cyan,
    pdftitle={Overleaf Example},
    pdfpagemode=FullScreen,
    }
\usepackage{tikz}
\newcommand\copyrightnotice[1]{
    \begin{tikzpicture}[remember picture,overlay]
    \node[anchor=south,yshift=10pt] at (current page.south) {\fbox{\parbox{\dimexpr\textwidth-\fboxsep-\fboxrule\relax}{#1}}};
    \end{tikzpicture}
}

\begin{document}
\title{Completeness of Datasets Documentation on ML/AI repositories: an Empirical Investigation}
%\title{An Empirical Investigation into the State of Datasets Documentation Practice in ML/AI repositories}
%
% \titlerunning{On the State of Datasets Documentation Practice}
\titlerunning{Completeness of Datasets Documentation on ML/AI repositories}
%\author{Anonymous Author(s)}
%\authorrunning{Anon.}
 \author{Marco Rondina\inst{1} \and
 Antonio Vetrò\inst{1} \and
 Juan Carlos De Martin\inst{1}}
 \institute{Politecnico di Torino, Corso Duca degli Abruzzi 24, 10129 Torino, Italy
 \email{\{marco.rondina,antonio.vetro,demartin\}@polito.it}}
 \authorrunning{M. Rondina et al.}
\maketitle              % typeset the header of the contribution
\begin{abstract}
% The abstract should briefly summarize the contents of the paper in 15--250 words.
ML/AI is the field of computer science and computer engineering that arguably received the most attention and funding over the last decade.
Data is the key element of ML/AI, so it is becoming increasingly important to ensure that users are fully aware of the quality of the datasets that they use, and of the process generating them, so that possible negative impacts on downstream effects can be tracked, analysed, and, where possible, mitigated. 
One of the tools that can be useful in this perspective is dataset documentation.

The aim of this work is to investigate the state of dataset documentation practices, measuring the completeness of the documentation of several popular datasets in ML/AI repositories.
We created a dataset documentation schema—the Documentation Test Sheet (\textsc{dts})—that identifies the information that should always be attached to a dataset (to ensure proper dataset choice and informed use), according to relevant studies in the literature.
We verified 100 popular datasets from four different repositories with the \textsc{dts} to investigate which information was present.

Overall, we observed a lack of relevant documentation, especially about the context of data collection and data processing, highlighting a paucity of transparency.

\end{abstract}
\section{Introduction and motivation}
\copyrightnotice{This preprint has not undergone peer review (when applicable) or any post-submission improvements or corrections. The Version of Record of this contribution is published in ``Progress in Artificial Intelligence. EPIA 2023. Lecture Notes in Computer Science(), vol 14115. Springer, Cham.'', and is available online at \href{https://doi.org/10.1007/978-3-031-49008-8_7}{https://doi.org/10.1007/978-3-031-49008-8\_7}}
Machine Learning / Artificial Intelligence (ML/AI) research made great strides in recent years, and its industrial applications became increasingly pervasive within society, automating organizational processes and decisions in several fields.
 
% Many issues related to fairness, transparency, and accountability in ML/AI systems are rooted in the data collection and in the data processing procedures, to the extent that proposals emerged for professionals specifically dedicated to these delicate stages of development \cite{joLessonsArchivesStrategies2020}.
% Every decision made during the workflow may contain implicit values and beliefs \cite{scheuermanDatasetsHavePolitics2021}, so tracking them can improve transparency \cite{afzalDataReadinessReport2021}.

% For example, in the context of a dataset with the purpose of training a facial recognition software, the way in which photos in the dataset are taken can have a significant impact on the results\cite{scheuermanDatasetsHavePolitics2021,wuResponsesCritiquesMachine2017}.
% In addition, the circumstances around data collection and data processing can undermine the ability of the dataset to universally abstract reality (although it is important to point out that probably it is impossible to standardize a classification about the world \cite{crawfordAtlasAI2021}).

Datasets are fundamental in the ML/AI ecosystem and many issues related to fairness, transparency, and accountability in ML/AI systems are rooted in the data collection and in the data processing procedures \cite{joLessonsArchivesStrategies2020}.
Every decision made during the workflow may contain implicit values and beliefs \cite{scheuermanDatasetsHavePolitics2021}, so tracking them can improve transparency \cite{afzalDataReadinessReport2021}.
The information accompanying them plays a very significant role in uncovering data issues \cite{boydDatasheetsDatasetsHelp2021}, in fostering reproducibility and auditability \cite{konigstorferSoftwareDocumentationNot2021}, in ensuring accountability \cite{hutchinsonAccountabilityMachineLearning2021}, users' trust \cite{arnoldFactSheetsIncreasingTrust2019}, and in avoiding \textit{data cascading} effects on the entire ML/AI pipeline \cite{sambasivanEveryoneWantsModel2021}. %and in organizing datasets \cite{halevyGoodsOrganizingGoogle2016}.
With documentation, it is possible to better understand the characteristics of the training data to at least partially mitigate the risks of downstream negative effects \cite{benderDangersStochasticParrots2021}.
Documentation production should be seen as an essential part of dataset production, as a place to disclose fundamental choices, in parallel with what is proposed to be documented in terms of models \cite{mitchellModelCardsModel2019,richardsMethodologyCreatingAI2020} or rankings \cite{yangNutritionalLabelRankings2018,zehlikeFairnessRankingSurvey2021}.
% While interesting proposals on post hoc documentation are catching on \cite{fabrisAlgorithmicFairnessDatasets2022,bandyAddressingDocumentationDebt2021}, this procedure is not always feasible, especially in case of very large datasets.

This study focuses on the dataset documentation state of practice.
The aim was to measure whether and how much relevant information about data collection and data processing procedures is present in the documentation of the most popular (and influential \cite{kochReducedReusedRecycled2021}) datasets. 
The research question which directed the design of the research is: \textbf{Which of the information, that should be transparent to dataset users, is present in the most popular datasets in ML/AI repositories?}
In order to answer this research question, we developed a test schema to measure the completeness of dataset documentation: Section \ref{sec:recommended-information-scheme} will describe the construction of the \textsc{dts}.
Subsequently, Section \ref{sec:study-design} describes the selection of repositories and datasets, according to a popularity proxy.
The results of the application of the \textsc{dts} are presented in Section \ref{sec:results}.
Finally, limitations (Section \ref{sec:limitations}), future work (Section \ref{sec:future-work}) and the conclusions (Section \ref{sec:conclusions}) are presented.
Furthermore, we provide additional materials in the online Appendix\footnote{Temporarily at \url{https://mega.nz/folder/ddAmXBpC\#sRITh3kS6i_9S7xa3-ECAg}.
The Appendix includes: the \textsc{dts} (A), a description of the provenance of the field of information composing it (B), the metadata of the selected datasets (C), the reading principles that guided the documentation investigation (D), the raw results (E) and additional tables and figures (F).} to enable external validation and enhance reproducibility.
%

%\section{Study of a recommended information schema obtained from related works}\label{sec:recommended-information-scheme}
\section{Documentation Test Sheet from related works}\label{sec:recommended-information-scheme}

We built a collection of recommended information that should be present in dataset documentation to ensure a proper choice of dataset and informed use.
The aim was to recognize, with a study of relevant work in the literature of dataset documentation schemas, which information is important to be present in dataset documentation to achieve transparency, accountability, and reproducibility.
The goal of this schema is to measure how complete a dataset documentation is: this property is the first necessary element to be scrutinized for enabling any further analysis on further quality dimensions of documentation (e.g., correctness).
We called this schema the \textit{Documentation Test Sheet} (\textsc{dts}).

\subsection{Fields of information}

% \begin{table}[t]
% \caption{Extract from the Documentation Test Sheet, Section 1 Motivation.}
% \label{tab:documentation-test-sheet}
% \begin{tabular}{|p{0.125\textwidth}|p{0.7\textwidth}|p{0.1\textwidth}|}
%     \hline \multicolumn{3}{|l|}{\footnotesize\textbf{Dataset:}} \\ \hline \multicolumn{1}{|p{0.125\textwidth}|}{\footnotesize\textbf{Test Field ID}} & \multicolumn{1}{p{0.75\textwidth}|}{\footnotesize\textbf{Test Field Name}} & \multicolumn{1}{p{0.12\textwidth}|}{\footnotesize\textbf{Presence Check}} \\ \hline

%     1.01 & \textit{Purpose for the dataset creation} &  \\ \hline
%     1.02 & \textit{Dataset creators} &  \\ \hline
%     1.03 & \textit{Dataset funders} &  \\ \hline
%     \multicolumn{2}{|r|}{\textbf{1} \textit{Motivation} \textit{Presence Average}} & \\ \hline 
% \end{tabular}
% \end{table}

The list of \textit{Test Fields} is largely based on \textit{Datasheets for Datasets}  \cite{gebruDatasheetsDatasets2021}, with some further insights from relevant documentation standardization proposals in the literature \cite{hollandDatasetNutritionLabel2018,benderDataStatementsNatural2018}.
We grouped the information into 6 sections, following the categorization presented in \textit{Datasheets for Datasets} (DfD): \textbf{1} \textit{Motivation}, \textbf{2} \textit{Composition}, \textbf{3} \textit{Collection processes}, \textbf{4} \textit{Data processing procedures}, \textbf{5} \textit{Uses}, \textbf{6} \textit{Maintenance}.
In addition to dataset metadata, some characteristics of the data were tracked in section \textbf{c} \textit{Characteristics}.
We discarded the \textit{Distribution} section because it proved inapplicable when testing documentation of datasets already published in public repositories.
% The table \ref{tab:documentation-test-sheet} shows the section \textbf{1} \textit{Motivation}, 
The full \textsc{dts} can be found in Appendix A, but the list of \textit{Test Fields} is also available in Table \ref{tab:fields-avg}.
Further details on the motivations behind this choice, and a description of the provenance of each information field, are reported in Appendix B.
One of the novelties of this work is the design of the individual \textit{Test Fields} as concepts expressed by few words to which it is easy to answer \say{yes} or \say{no}, depending on the presence or absence of the related information in the documentation under analysis.
Some fields from the related work were collapsed in a few \textit{Test Fields} for the sake of brevity and ease of application.
We designed the \textsc{dts} to be generalizable as possible to any type of documentation, so that it can be used for datasets pertaining to different areas of ML/AI.

\subsection{Measurement}\label{subsec:measurement}
The other core elements of the \textsc{dts} are the \textit{Presence Check Values} and the \textit{Presence Averages}.
During the analysis of the documentation, each \textit{Test Field} is associated with a value indicating the presence or the absence of the represented information.
Specifically, the \textit{Presence Check Value} can take on one of the following three possible values:
\begin{itemize}
    \item 1: it is possible to retrieve the information represented by the \textit{Test Field};
    \item 0: it is not possible to retrieve the information represented by the \textit{Test Field};
    \item NA: the information represented by the \textit{Test Field} does not apply to dataset;
\end{itemize}

The \textit{Presence Average} represents the completeness measure of the \textsc{dts}. 
It is obtained by averaging the \textit{Presence Check Values} of the group of \textit{Test Fields} under analysis such as dataset (\textit{Dataset Presence Average}), section (\textit{Section Presence Average}), and field (among different datasets, \textit{Field Presence Average}). 
%As part of the discussion, there will also be an analysis of the data distributions in order to have a complete overview of the results.

\section{Study design}\label{sec:study-design}
%We structured the analysis of dataset documentation to understand how much of the documentation under investigation was complete and which sections of information were most complete or lacking. 
One of the novel elements of this study concerns the analysis of the information found in the very same place where the data can be accessed, instead of selecting datasets from a corpus of academic papers \cite{pengMitigatingDatasetHarms2021,geigerGarbageGarbageOut2020}.
The documentation in the public repository provides information about how the dataset is actually used in practice.
Since the purpose of a scholarly article is different from that of a repository, some information may have no reason for inclusion in the article, and vice versa.
For this reason, the documentation being analysed is the documentation web page where data can be downloaded. 
Given this design choice, we first selected the repositories under analysis, as described in section \ref{subsec:repo-under-analysis}.
In the second step, we collected the metadata useful to perform the dataset selection, as described in section \ref{subsec:datasets-selection}.
We focused on the most popular datasets, as seen in other work \cite{fabrisAlgorithmicFairnessDatasets2022}. This is because these datasets are the most influential ones, and therefore studying their documentation is an important step in the path towards a deeper understanding of common documentation practices.
We then collected data on the presence of information in the documentation \footnote{For reasons of space, summary tables with raw data are presented in Appendix E}.

\subsection{Repositories under analysis}\label{subsec:repo-under-analysis}
The choice of repository is a relevant decision in this study, due to the design decision to analyse the online documentation present in the same place where the data are hosted. 
Indeed, different repositories have different documentation and metadata schemas.
Therefore, we decided to select more than one repository to avoid obtaining too specific results. 
We selected four well-known and commonly used repositories to capture different practices in the ML/AI community.
The criteria used for the choice were: free access; the presence of popularity proxies among the metadata; the presence of hundreds of datasets. 
% we selected three generalist repositories\footnote{The term \say{generalist repository} refers to repositories not focused on a single sector of artificial intelligence applications (e.g. computer vision)} and one repository related to the world of Natural Language Processing (NLP)\footnote{In addition, please consider that this type of sector is growing thanks to deep learning \cite{otterSurveyUsagesDeep2021} and applications from this sector require special attention to ethical issues \cite{benderDataStatementsNatural2018}}, because of the presence of \textit{Data Statements for Natural Language Processing} \cite{benderDataStatementsNatural2018} as source of the \textit{Documentation Test Sheet}. 
We consulted the Wikipedia \textit{List of portals suitable for multiple types of ML applications}\footnote{\href{https://en.wikipedia.org/wiki/List\_of\_datasets\_for\_machine-learning\_research}{https://en.wikipedia.org/wiki/List\_of\_datasets\_for\_machine-learning\_research}} and, as a result, the following repositories have been selected:  \href{https://huggingface.co/datasets}{Hugging Face}\footnote{Since data collection (see Appendix E.1), the repository's dataset count has increased nearly tenfold, and now includes additional sectors like audio and computer vision.} (\textsc{hug}), \href{https://www.kaggle.com/datasets}{Kaggle} (\textsc{kag}), \href{https://www.openml.org/search?type=data}{OpenML} (\textsc{oml}) and \href{https://archive-beta.ics.uci.edu/ml/datasets}{UC Irvine Machine Learning Repository} (\textsc{uci}).

\subsection{Datasets selection}\label{subsec:datasets-selection}
To guarantee the feasibility of the research, it was also necessary to limit the number of datasets for each repository.
For this reason, we selected 25 datasets from each repository, for a total of 100 datasets to be examined.
We decided to focus on the concept of \textit{popularity}, so that we could analyse some of the most used and influential datasets: where available, the number of \textit{downloads} was identified as the best proxy; where not available, number of \textit{views} was identified as a good alternative.
% To represent this concept, we needed a parameter that should be: transparent, i.e. the way how it is calculated should be clearly documented; common, i.e. the value should apply to different repositories; sortable: i.e. it is possible to sort datasets based on that value.
The resulting metrics used are: \textsc{hug}, number of downloads (APIs); \textsc{kag}, platform upvotes and then the number of downloads (APIs)\footnote{Due to the unavailability of direct download count APIs, datasets were sorted by upvotes via APIs and then sorted by download count, as presented in the results.}; \textsc{oml}, number of downloads (web scraping); \textsc{uci}, number of views (web scraping).
% \footnote{The number of views is a transparent parameter, which is able to reflect the popularity of the datasets, and it is reasonable to think that it is a good proxy for the number of downloads}
We eliminated any duplicates within the same or different repositories (the comparison of information about the same dataset in different repositories was not a central aim of this research).
As a selection criterion between duplicates, we used the highest \say{popularity}. 
In the case of two datasets at the same ranking position, we eventually observed whether one of them was the primary source of the other.
The full list of selected datasets, together with the date of data collection, can be found in Appendix C.
The principles that guided the reading of the documentation are presented in the Online Appendix D.

\section{Results and discussion}\label{sec:results}
%This chapter presents the results obtained by the documentation analysis, with further discussion.
In this section, we jointly present the results and their discussion for each of the following levels: dataset (\ref{sec:results-datasets}),  \textsc{dts} section (\ref{sec:results-sections}), \textit{Test Fields} (\ref{sec:results-fields}).
Additional information on the distribution and dispersion of values is included in Appendix F.

\subsection{Datasets level}\label{sec:results-datasets}

\begin{figure}[t]
    \centering
    \includegraphics[width=1\linewidth]{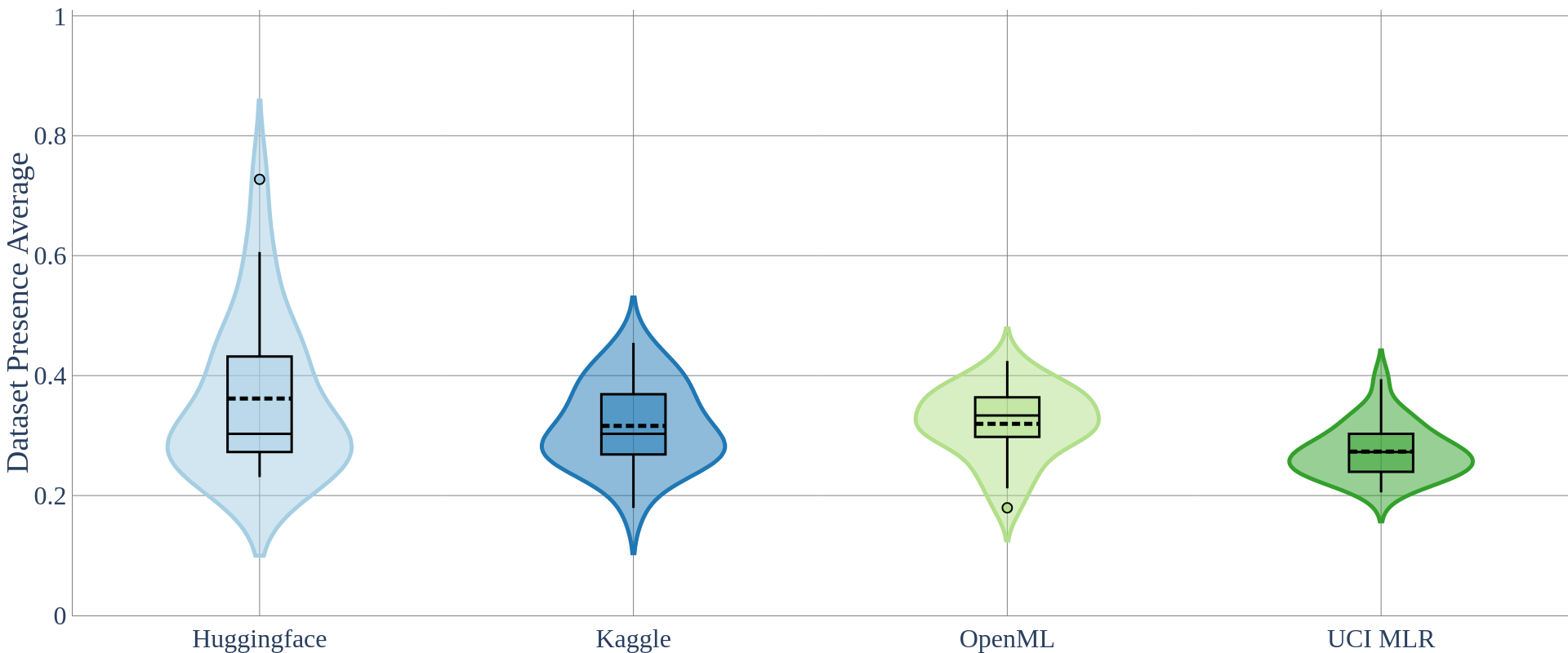}
    \caption[Distribution of Dataset Presence Averages grouped by repository]{Distribution of Dataset Presence Averages grouped by repository.}
    \label{fig:2_gbyRepositoryVB}
    % \Description{Distribution of Dataset Presence Averages grouped by repository. For each repository, the violin plot shows the frequency of each Dataset Presence Average value. Huggingface obtained the best results with an oulier near 0,70 and a mean value equal to 0,36. The mean values of the three others repositories range from 0,27 to 0,31. OpenML shows an outliers with value below 0,20.}
\end{figure}

% This section will analyse the data from the perspective of datasets, first from a general point of view, then on the basis of datasets characteristics. 
The dataset with the most comprehensive documentation is \textbf{hug16}, the \href{https://huggingface.co/datasets/cnn_dailymail}{\textit{cnn dailymail}} from Hugging Face (\textsc{hug}).
It contains over 300k unique news articles written by journalists.
Its \textit{Dataset Card}\footnote{\say{Dataset Card} is the name of the documentation schema attached to each dataset in the Hugging Face repository (\href{https://huggingface.co/docs/hub/datasets-cards}{https://huggingface.co/docs/hub/datasets-cards}).} was comprehensive in all the different sections, and it can be considered a positive reference point from the point of view of documentation practice.

Figure \ref{fig:2_gbyRepositoryVB} shows that overall very few datasets achieve more than 50\% completeness, and variation between repositories is small.
The selected datasets from \textsc{hug} have the highest mean of the \textit{Dataset Presence Average} distribution,  while the ones from \textsc{uci} have the lowest mean of the \textit{Dataset Presence Average} distribution. 
One of the contributing factors to this result is that the three most complete documentations belong to \textsc{hug} datasets.

\input{tables/datasets-characteristics}

In Table \ref{tab:char} it is possible to observe specific characteristics of the datasets: most datasets do not contain personal data, have an explicit target variable, and are not a sample or reduction of a larger set.
All datasets updated after 1 January 2021 were considered \say{Recently updated}: all datasets from \textsc{hug} and five datasets from \textsc{kag} are recently updated in terms of data or documentation, while all the \textsc{oml} and \textsc{uci} datasets have not been updated in this timeframe.
% These results may provide another explanation for the better documentation quality of the most popular datasets in \textsc{hug}.

Additional statistics from further segmentation of results at the dataset level can be found in Appendix F.
% It i s possible to observe how the \textit{Presence Average} of the whole dataset varies on the basis of its characteristics.
%Figure 1 in the Appendix F presents the variation of the mean amount of information accompanying the dataset according to the presence (or absence) of people-related data.
%This Figure shows that datasets containing people-related data present on average less complete documentation, but the differences are too small to speculate on this.

\subsection{Sections level}\label{sec:results-sections}
\begin{figure}[th]
    \centering
    \includegraphics[width=1\textwidth]{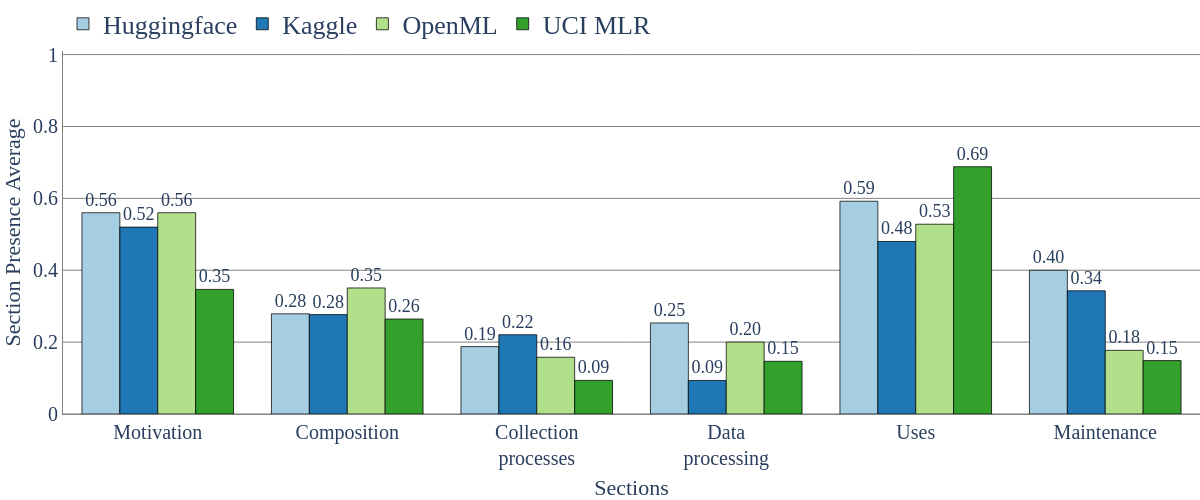}
    \caption[Sections Presence Average]{\textit{Section Presence Averages}.}
    \label{fig:3_gbySectionRepository}
    % \Description{Section Presence Averages. The bar plot shows the mean value of each repository for all the sections of the Documentation Test Sheet. Each value represents the average amount of information founded in the given section of the given repository. The section with the best results is the Uses one, with Section Presence Averages values equal to 0.59, 0.48, 0.53 and 0.69 (respectivly for Huggingface, Kaggle, OpenML and \textsc{uci} MLR).}
\end{figure}

% A further level of analysis is the analysis by category of information, i.e.\ sections.
As can be seen in Figure \ref{fig:3_gbySectionRepository}, the \textit{Uses} section was the most complete one, followed by the \textit{Motivation} section. Sections \textit{Collection processes} and \textit{Data processing procedures} had the lowest values of \textit{Section Presence Average}. Additionally, we observed that the results of the \textit{Maintenance} section are very different between repositories.

%Overall, the analysis shows that documentation available on the repositories is mostly related to data usages.  %qui commento 1O4
These results suggest that the documentation of public datasets is currently utilisation-oriented, with less attention to the previous stages of the dataset construction pipeline.
This aspect is also correlated with the high \textit{Section Presence Average} of the \textit{Motivation} section: the purpose of the dataset often encapsulates the meaning of why the data within it should be used. 
The low completeness of the \textit{Composition}, \textit{Collection processes} and \textit{Data processing procedures} sections suggests that either little effort is devoted to describing the early stages of the dataset construction phase.
Frequently, there is no information at all about these delicate phases.
The failure to take into account these contextual aspects can lead to various problems in the models trained on such undocumented data \cite{sambasivanEveryoneWantsModel2021}. 
Recent work devoted to partially automating the documentation process could help users to easily complete the \textit{Composition} section \cite{schramowskiCanMachinesHelp2022,sunMithraLabelFlexibleDataset2019,petersenDataMaidYourAssistant2019}.
% ,arslanHowAutomaticallyDocument2019
Finally, the \textit{Maintenance} results, although very variable between repositories, confirmed recent studies in the literature about the  opportunities to improve documentation in datasets repositories and the lack of attention paid to what happens after the dataset is published \cite{corryProblemZombieDatasets2021,thylstrupEthicsPoliticsData2022}.
As for the other aggregation level, the distributions of measurements are provided in Appendix F.

%the peculiarities and possible dangers of the data contained in the dataset.  

% OLD
% Figure \ref{fig:3_gbySectionRepository} makes it possible to see at a glance which parts of the documentation under analysis receive the most attention.
% Data shows that dataset publishers pay the majority of their focus on information related to how to use the data contained within them. 
% This result testifies the extent to which the documentation production of public datasets is currently utilisation-oriented, without paying too much attention to other relevant aspects in previous stages of the dataset construction.
% It is indeed very difficult to find information concerning all the various data \textit{Collection processes} and \textit{Data processing procedures}: extremely delicate phases, in which the failure to take certain contextual aspects into account can lead to various problems in the models trained on these data \cite{sambasivanEveryoneWantsModel2021}.
% The high result of the \textit{Motivation} section can also be partially justified by a greater focus on purely \say{technical} aspects: the purpose of the dataset often encapsulates the meaning of why the data within it should be used.
% Finally, the \textit{Maintenance} shows the lack of attention paid to what happens after the dataset is published. This finding is consistent with the fact that this is an open research question \cite{corryProblemZombieDatasets2021,thylstrupEthicsPoliticsData2022}.

\subsection{Test Fields level}\label{sec:results-fields}
%The last level of investigation is the individual fields of information. 
%There are certain fields whose frequent presence (or, conversely, whose frequent absence) was a common feature of all the repositories.
Table \ref{tab:fields-avg} shows the \textit{Presence Average} value of each information field, globally and by the repository.
Results show that certain documentation fields are very commonly used, such as the \textbf{2.01} \textit{Description of the instances} (0,92), \textbf{2.02} \textit{Number of the instances} (0,90) and
\textbf{5.01} \textit{Description of the tasks in which the dataset has already been used and their results} (0,95).
In many cases, the high level of completeness could be explained by the ability of the repository's metadata structure to promote the presence of a particular piece of information.
Indeed, the information represented by these fields was very much present in repositories that structurally expose this information in the metadata schema of the repository.
Conversely, it was almost completely absent in repositories that do not include such information in their metadata schema.
Some examples are: \textbf{2.11} \textit{Statistics} (\textsc{hug} 0,00; \textsc{kag} 1,00; \textsc{oml} 1,00; \textsc{uci} 1,00), \textbf{5.04} \textit{Repository that links to papers or system that use the datasets} (\textsc{hug} 0,92; \textsc{kag} 0,00; \textsc{oml} 0,00; \textsc{uci} 1,00), \textbf{5.05} \textit{Description of license and terms of use} (\textsc{hug} 0,48; \textsc{kag} 0,68; \textsc{oml} 1,00; \textsc{uci} 1,00).
This highlights the role played by repository hosts, who have the potential to trigger virtuous documentation practices.

Looking further into the results, additional observations can be outlined.
In the \textit{Motivation} section, on the one hand, it was very common to find information about the authors (the \say{resource creators}), while on the other hand, it was rare to find details about who funded the creation of the dataset, important information for achieving accountability. %the \say{resource creators}following the taxonomy proposed by McMillan-Major et al. \cite{mcmillan-majorReusableTemplatesGuides2021}
Within the \textit{Composition} section, basic information such as the description or number of instances was usually present. 
On the contrary, information about data confidentiality and dangerousness was usually absent.  
%Furthermore, in order to reinforce the potential contribution that the repository scheme can make to improving completeness, graphical and tabular statistics, pair plots, probabilistic models and ground truth correlations are data insights that could easily be generated automatically. %The fact that the repositories hosting the datasets allow for easy integration of this content may have an impact on the availability of such information. 
The analysis of information related to \textit{Collection processes} pointed out, in a context of a general scarcity of details, the near total absence of specifics about ethical review processes and about analysis of potential impacts of dataset uses.
With regard to the \textit{Data processing procedures}, we observed that the \say{Dataset Card} in HUG favoured the presence of (at least) some useful tags to obtain indications on the workers involved in these procedures. 
%On the contrary, no similar hints are provided in the other repositories investigated.
As already mentioned above, in terms of \textit{Uses}, much attention on the part of dataset creators is paid to the description of the previous usage of the dataset and to the description of the recommended uses.
The same cannot be said for non-recommended uses: only the documentation of a couple of \textsc{hug} datasets contained this information.
Surprisingly, although it was common to find details on the subject that supports or manages the dataset, the contact of the owner was rarely present.
Furthermore, in terms of \textit{Maintenance}, the DOI was quite rare and no information on the management of older dataset versions could be retrieved.

\input{tables/field-presence-averages}

\section{Threats to validity and limitations}\label{sec:limitations}
%limitazioni
One of the main limitations of this research is the non-scalability of the proposed procedure, which was primarily based on manual inspection of dataset documentation: alignment of repositories metadata with the documentation fields proposed in the literature and included in the \textsc{dts} was very poor.
% Moreover, the number and the types of the selected repositories were focused by design on some artificial intelligence sectors, neglecting other relevant ones, such as computer vision.

The choice of repositories may have influenced the final result.
However, by focusing on some of the most prominent repositories and the most popular datasets in each repository, we were able to analyse the documentation of influential datasets.
The dataset selection criteria - \textit{popularity} - was implemented slightly differently to the different repositories, due to differences in the metadata schemas: however, the number of downloads was present in three out of four repositories, and for the remaining one we selected the most reasonable and available proxy (visualizations).
In addition, popularity is prone to be a proxy of longevity: this criterion may have introduced a selection bias, favouring datasets from a time when documentation was less important or emphasized.
On the contrary, the lack of documentation updates on such datasets reinforces the findings of this study, i.e. poor attention/availability on dataset documentation.

Despite the fact that considerable effort has been made to make the data collection as accurate and standardized as possible, the study design, strongly based on human reading and interpretation of documentation texts, is inherently prone to the risk of interpretation errors.
We controlled this threat by providing the reading principles in Appendix D.

%A further limiting aspect is represented by the fact that the procedures exposed in this research evaluated only a single aspect of dataset documentation - \textit{completeness} - leaving out other quality aspects (correctness, relevance, etc).

Finally, due to lack of resources, the DTS was not tested for consistency and validation with target users: however, the information fields were all derived from documentation schemes already available in the academic literature.

\section{Future work}\label{sec:future-work}
%future work
The obtained results and the limitations highlighted provide insights and suggestions on possible expansions of this research. 
The first hypothesis of future work is related to increasing the number of datasets and repositories under investigation.
% This would provide a more complete view of the state of the art and could allow us to investigate whether there is any form of correlation between the popularity of a dataset and the completeness of the documentation related to it (correlation not revealed by the data obtained through the datasets selected for this work).
Moreover, a complementary analysis of a selection of recent datasets could tell us if the growing awareness of data curation is bringing some results in common practice.
Quantitative expansions of the research could be put in place starting with investigations into the feasibility of an automatic system capable of controlling the presence of information: this possibility, however, is fully dependent on the evolution of the repository, and actions made possible by dataset hosts to standardize documentation and make it machine-readable.

From the qualitative point of view, it might be possible to expand the \textsc{dts} to measure other aspects of documentation quality.
For example, comparing the information found in the repositories with the information retrieved from academic articles using those datasets could reveal further insights to understand documentation practices, reduce documentation debt and possibly integrate it with additional aspects (e.g., \say{sparsity} \cite{fabrisAlgorithmicFairnessDatasets2022}, dataset quality). %Moreover, this possible research could be useful in identifying the most popular and at the same time most problematic datasets from a documentation point of view. 
%Retrospective documentation work could be carried out on those datasets in order to reduce documentation debt, as shown in \cite{fabrisAlgorithmicFairnessDatasets2022,bandyAddressingDocumentationDebt2021,garbinStructuredDatasetDocumentation2021}.
%it could be interesting to couple the results obtained with the documentation analysis with some investigation on the data itself, taking into account some aspects of data quality.
Finally, a test with target users that also explores the differences between different types of dataset users could be useful for prioritizing \textsc{dts} Test Fields according to possible users and uses.

\section{Conclusions}\label{sec:conclusions}
We empirically investigated the state of documentation practice in the most popular datasets in the ML/AI community.
%by measuring how much of the relevant information is present in the documentation of
A set of information that should always be clear to the users of the datasets, in order to achieve transparency and accountability, was adapted into a \textit{Documentation Test Sheet} (\textsc{dts}) able to measure the completeness of documentation.
The \textsc{dts} was applied to 100 dataset documentations from Hugging Face, Kaggle, OpenML and UC Irvine MLR repositories.
%In order to address the research question \textbf{Which of the information, that should be transparent to dataset users, is present in the most popular datasets in ML/AI repositories?}, manual inspection has been performed on 100 dataset documentation (from Hugging Face, Kaggle, OpenML and \textsc{uci} ML repositories), focusing on the information available in the very same place where data can be accessed.

This investigation brought out some relevant results about the state of practice of documentation of datasets manufacturing. First, it emerged that information related to how to use the dataset was the most present.
On the contrary, maintenance over time or processes behind the data generation were very poorly documented.
In general, a lack of relevant information was observed, highlighting a paucity of transparency.
All these observations are even more relevant when considering that the analysis was restricted to some of the most popular and well-known datasets.
Finally, the potential of repositories to help curators of datasets to produce better documentation emerged. 
% Indeed, when repositories provided a more comprehensive documentation schema, the resulting documentation was more complete.

Altogether, these results let us hypothesize that efforts of the ML/AI community in devoting more attention to the dataset documentation process are necessary.
These efforts might enable the reuse of datasets in a way that is more aware of the choices, assumptions, limitations and other aspects of their creation, and ultimately facilitating human-respectful ML/AI innovations.
The proposed \textsc{dts} can be an easy-to-use tool in the hands of dataset creators, maintainers, and hosts to move a further step in this direction. 

%The actual practice of industry practitioners is slightly different from the scenario outlined by fair ML/AI research literature \cite{holsteinImprovingFairnessMachine2019}.
%There are no purely technical aspects, and every \say{technical} choice that led to the construction of a given model hides behind it a set of ethical considerations, regardless of whether the context is to be taken into account or not.
%The recommended path should be supported by the investigation and experimentation of techniques to fully integrate documentation models and processes into the ML/AI pipeline. 

%Moreover, it can serve as a guideline for dataset creators, helping them to improve their documentation so that dataset consumers can verify the underlying choices and assumptions.
% \vfill
\noindent\textbf{Acknowledgements.}
This study was carried out within the FAIR - Future Artificial Intelligence Research and received funding from the European Union Next-GenerationEU (PIANO NAZIONALE DI RIPRESA E RESILIENZA (PNRR) – MISSIONE 4 COMPONENTE 2, INVESTIMENTO 1.3 – D.D. 1555 11/10/2022, PE00000013). This manuscript reflects only the authors’ views and opinions, neither the European Union nor the European Commission can be considered responsible for them.
\bibliographystyle{splncs04}
\bibliography{01_dts}




% LNCS Homepage, \url{http://www.springer.com/lncs}. Last accessed 4
% \end{thebibliography}
\end{document}

%% file: tables/datasets-characteristics.tex
\begin{table}[t]
    \caption[Characteristics of the 100 selected datasets among different repositories]{Characteristics of the 100 selected datasets (25 for each repository).}
    \label{tab:char}
    % \centering
    \begin{tabular}{|p{0.2\textwidth}|p{0.1\textwidth}|p{0.2\textwidth}|p{0.3\textwidth}| p{0.12\textwidth}|}
        \hline
        \multicolumn{1}{|p{0.2\textwidth}|}{\footnotesize{\textbf{Repository}}} &
        \multicolumn{1}{|p{0.1\textwidth}|}{\footnotesize{\textbf{Data is people related}}} &
        \multicolumn{1}{|p{0.2\textwidth}|}{\footnotesize{\textbf{Presence of explicit target variable}}} &
        \multicolumn{1}{|p{0.3\textwidth}|}{\footnotesize{\textbf{Dataset is a sample or a reduction of a larger set}}} &
        \multicolumn{1}{|p{0.12\textwidth}|}{\footnotesize{\textbf{Recently\newline updated}}} \\
        % \multicolumn{1}{m{2cm}}{\footnotesize\textbf{Repository}} & \multicolumn{1}{m{2.5cm}}{\footnotesize\textbf{Data is people-related}} & \multicolumn{1}{m{2.5cm}}{\footnotesize\textbf{Presence of explicit target variable}} & \multicolumn{1}{m{2.5cm}}{\footnotesize\textbf{Dataset is a sample or a reduction of a larger set}} & \multicolumn{1}{m{2cm}}{\footnotesize\textbf{Recently updated}} \\
        \hline
        Hugging Face   & 04 & 21 & 01 & 25  \\ 
        Kaggle         & 12 & 08 & 02 & 05  \\  
        OpenML         & 11 & 25 & 07 & 00  \\   
        UCI MLR        & 11 & 22 & 04 & 00  \\
        \hline
        \footnotesize{\textbf{Total}}          & 38 & 76 & 14 & 30  \\
        \hline
    \end{tabular}
\end{table}

%% file: tables/field-presence-averages.tex
\begin{longtable}[t]{lp{6.82cm}lllll}
    \caption[Fields Presence Averages]{\textit{Field Presence Averages}: overall and for each repository.\\}
    \label{tab:fields-avg} \\

    \hline  \footnotesize\textbf{ID} & \footnotesize\textbf{Field description} & \footnotesize\textbf{Tot} & \footnotesize\textbf{HUG} & \footnotesize\textbf{KAG} & \footnotesize\textbf{OML} & \footnotesize\textbf{UCI} \\ \hline
    \endfirsthead

    \multicolumn{6}{c}%
    {\tablename\ \thetable{} -- continued from previous page.} \\
    \hline  \footnotesize\textbf{ID} & \footnotesize\textbf{Field description} & \footnotesize\textbf{Tot} & \footnotesize\textbf{HUG} & \footnotesize\textbf{KAG} & \footnotesize\textbf{OML} & \footnotesize\textbf{UCI} \\ \hline
    \endhead

    \hline \multicolumn{6}{r}{{Continue on next page.}} 
    \endfoot

    \hline
    \endlastfoot
    \hline

    1.01 & \textit{Purpose for the dataset creation} & 0,57 & 0,64 & 0,52 & 0,68 & 0,44 \\
    1.02 & \textit{Dataset creators} & 0,86 & 0,88 & 0,96 & 1,00 & 0,60 \\
    1.03 & \textit{Dataset funders} & 0,06 & 0,16 & 0,08 & 0,00 & 0,00 \\ \hline
    2.01 & \textit{Description of the instances} & 0,92 & 1,00 & 1,00 & 0,80 & 0,88 \\
    2.02 & \textit{Number of the instances} & 0,90 & 0,92 & 0,72 & 1,00 & 0,96 \\
    2.03 & \textit{Information about missing values} & 0,50 & 0,00 & 0,12 & 1,00 & 0,88 \\
    2.04 & \textit{Recommended data splits} & 0,31 & 0,92 & 0,08 & 0,12 & 0,12 \\
    2.05 & \textit{Description of errors, noise or redundancies} & 0,13 & 0,00 & 0,16 & 0,08 & 0,28 \\
    2.06 & \textit{Information about data confidentiality} & 0,04 & 0,08 & 0,08 & 0,00 & 0,00 \\
    2.07 & \textit{Information about possible data dangerousness (offensive, insulting, threatening or cause anxiety) or biases} & 0,03 & 0,12 & 0,00 & 0,00 & 0,00 \\
    2.08 & \textit{Information about people involved in data production and their compensation (if people related)} & 0,43 & 0,25 & 0,42 & 0,64 & 0,31 \\
    2.09 & \textit{Description of identifiability for individuals or subpopulations (if people related)} & 0,15 & 0,50 & 0,17 & 0,09 & 0,08 \\
    2.10 & \textit{Description of  data sensitivity (if people related)} & 0,03 & 0,25 & 0,00 & 0,00 & 0,00 \\
    2.11 & \textit{Statistics} & 0,50 & 0,00 & 1,00 & 1,00 & 0,00 \\
    2.12 & \textit{Pair plots} & 0,00 & 0,00 & 0,00 & 0,00 & 0,00 \\
    2.13 & \textit{Probabilistic model} & 0,00 & 0,00 & 0,00 & 0,00 & 0,00 \\
    2.14 & \textit{Ground truth correlations} & 0,00 & 0,00 & 0,00 & 0,00 & 0,00 \\ \hline
    3.01 & \textit{Description of instances acquisition and data collection processes} & 0,53 & 0,52 & 0,60 & 0,64 & 0,36 \\
    3.02 & \textit{Information about people involved in the data collection process and their compensation} & 0,08 & 0,16 & 0,12 & 0,00 & 0,04 \\
    3.03 & \textit{Time frame of data collection} & 0,19 & 0,04 & 0,48 & 0,12 & 0,12 \\
    3.04 & \textit{Information about ethical review processes} & 0,01 & 0,04 & 0,00 & 0,00 & 0,00 \\
    3.05 & \textit{Information on individuals' knowledge of data collection (if people related)} & 0,05 & 0,25 & 0,00 & 0,09 & 0,00 \\
    3.06 & \textit{Information on individuals' consent for data collection (if people related)} & 0,05 & 0,25 & 0,00 & 0,09 & 0,00 \\
    3.07 & \textit{Analysis of potential impacts of the dataset and its use on data subjects} & 0,00 & 0,00 & 0,00 & 0,00 & 0,00 \\ \hline
    4.01 & \textit{Description of sampling, preprocessing, cleaning, labelling procedures} & 0,39 & 0,32 & 0,24 & 0,56 & 0,44 \\
    4.02 & \textit{Information about people involved in the data sampling, preprocessing, cleaning, labelling procedures and their compensation} & 0,11 & 0,44 & 0,00 & 0,00 & 0,00 \\
    4.03 & \textit{Description of others possible sampling, preprocessing, cleaning, labelling procedures} & 0,02 & 0,00 & 0,04 & 0,04 & 0,00 \\ \hline
    5.01 & \textit{Description of the tasks in which the dataset has already been used and their results} & 0,95 & 0,92 & 1,00 & 1,00 & 0,88 \\
    5.02 & \textit{Description of recommended uses or tasks} & 0,62 & 0,56 & 0,72 & 0,64 & 0,56 \\
    5.03 & \textit{Description of not recommended uses} & 0,02 & 0,08 & 0,00 & 0,00 & 0,00 \\
    5.04 & \textit{Repository that links to papers or system that use the datasets} & 0,48 & 0,92 & 0,00 & 0,00 & 1,00 \\
    5.05 & \textit{Description of license and terms of use} & 0,79 & 0,48 & 0,68 & 1,00 & 1,00 \\ \hline
    6.01 & \textit{Information about subject supporting, hosting, maintaining the dataset} & 0,84 & 0,36 & 1,00 & 1,00 & 1,00 \\
    6.02 & \textit{Contact of the owner} & 0,30 & 0,20 & 0,80 & 0,16 & 0,04 \\
    6.03 & \textit{DOI} & 0,09 & 0,24 & 0,04 & 0,08 & 0,00 \\
    6.04 & \textit{Erratum} & 0,00 & 0,00 & 0,00 & 0,00 & 0,00 \\
    6.05 & \textit{Information about dataset updates} & 0,38 & 1,00 & 0,52 & 0,00 & 0,00 \\
    6.06 & \textit{Information about management of older dataset versions} & 0,00 & 0,00 & 0,00 & 0,00 & 0,00 \\
    6.07 & \textit{Information about the mechanism to extend, augment, build on, contribute to the dataset} & 0,26 & 1,00 & 0,04 & 0,00 & 0,00 \\ \hline
    c.01 & \textit{Data is people related} & 0,38 & 0,16 & 0,48 & 0,44 & 0,44 \\
    c.02 & \textit{Presence of explicit target variable} & 0,76 & 0,84 & 0,32 & 1,00 & 0,88 \\
    c.03 & \textit{Dataset is a sample or a reduction of a larger set} & 0,14 & 0,04 & 0,08 & 0,28 & 0,16 \\
    c.04 & \textit{Recently updated} & 0,30 & 1,00 & 0,20 & 0,00 & 0,00 \\ \hline
\end{longtable}